# Can neutrinos from neutron star mergers power $\gamma$-ray bursts ?

**H.-Th. Janka and M. Ruffert**

Max-Planck-Institut für Astrophysik, Karl-Schwarzschild-Str. 1, Postfach 1523, 85740 Garching, Germany

**Abstract.** We perform three-dimensional hydrodynamical simulations of the coalescence of binary neutron stars. We include the emission and backreaction of gravitational waves into the Newtonian "Piecewise Parabolic Method". The use of the physical equation of state (EOS) of Lattimer & Swesty (1991) allows us to take into account the production of neutrinos. We evaluate our models for the efficiency of $\nu\bar{\nu}$ annihilation in the surroundings of the coalescing neutron stars. The corresponding energy deposition prior to and during merging turns out to be 2–3 orders of magnitude too small to power a typical $\gamma$-ray burst with an energy output of $\sim (10^{51}/4\pi)$ erg/sterad at cosmological distances. Analytical estimates of the subsequent evolution of the disk which possibly surrounds the central black hole show that even under the most favorable conditions the energy provided by $\nu\bar{\nu} \to e^-e^+ \to \gamma\gamma$ falls short by at least an order of magnitude.

## 1. Introduction

The distribution of 1122 observed $\gamma$-ray bursts as reported by the BATSE 3B Catalog is isotropic over the sky and non-uniform in distance (Meegan et al. 1995). One of the currently most popular models for the origin of the $\gamma$ emission is the merger of two neutron stars (NS-NS) or of a neutron star and a black hole (NS-BH) at cosmological distances (Paczyński 1986, Goodman 1986, Eichler et al. 1989, Narayan et al. 1991, Paczyński 1991). These would occur at a rate of $10^{-4}$–$10^{-5}$ yr$^{-1}$ per galaxy and thus with a sufficiently large frequency to explain the observed burst rate (Phinney 1991, Narayan et al. 1991). Only about 1% of the several $10^{53}$ erg of energy released in the merging would be enough to power a burst, and the compactness of the source might account for the variations in the $\gamma$-ray flux on timescales as short as 1 ms.

The non-thermal $\gamma$-rays are supposed to be produced by a fireball of relativistic $e^-e^+$ pairs that are created by the annihilation of $\nu\bar{\nu}$ pairs in the vicinity of the hot, merged object (Goodman et al. 1987). However, the large $\nu$ fluxes also drive a baryonic wind by energy deposition near the neutrinosphere (Duncan et al. 1986, Woosley & Baron 1992) which "pollutes" the cloud of $\gamma$'s and $e^-e^+$ pairs. The baryonic load prevents relativistic expansion and keeps the flow optically thick, leading to a degradation of the $\gamma$-rays. Thus the energy release will be mainly kinetic rather than in photons (Paczyński 1990). If spherical symmetry is broken, as in the case of the merger scenario where a rapidly rotating disk can be formed and material is pulled away from the rotation axis by centrifugal forces, then a baryon-free funnel along the rotation axis may allow relativistic beams of $\gamma$'s and $e^-e^+$ to escape (Mészáros & Rees 1992a, Mochkovitch et al. 1993, Woosley 1993). Annihilation of $\nu\bar{\nu}$ along the system axis might lead to matter ejection with Lorentz factors of $\Gamma \sim 10^2$–$10^3$ (Mochkovitch et al. 1995) which are in the right range to enable copious $\gamma$ production during shock interaction with ambient interstellar gas (Rees & Mészáros 1992).

In this Letter we present 3D hydrodynamical simulations of the merging of binary NSs using the physical EOS of Lattimer & Swesty (1991) and including the effects of the backreaction of gravitational waves on the hydrodynamic flow (see Ruffert et al. 1996). The loss of lepton number and energy by $\nu$'s is simulated with an elaborate leakage scheme that was calibrated using diffusion results. It takes into account the emission of all types of $\nu$'s by thermal processes and of $\nu_e$ and $\bar{\nu}_e$ also by the $\beta$-processes. Neutrino opacities and diffusion timescales are calculated from $\nu$-$n,p$ scattering and the inverse $\beta$-reactions (Ruffert et al. 1996). We evaluate our models for the efficiency of $\nu\bar{\nu}$ annihilation and investigate the question whether the corresponding energy deposition is powerful enough to account for the energetics of burst events at cosmological distances. That is the crucial question when associating $\gamma$-ray bursts with NS-NS or NS-BH mergers, independent of the detailed, very complex processes which lead to the emission of observed $\gamma$-rays.

## 2. Thermal evolution and neutrino emission

*Pre-merging evolution.* Kochanek (1992) showed that unless the volume-averaged shear viscosity is larger than $\eta \gtrsim 10^{29}$ g cm$^{-1}$s$^{-1}$ the structure and evolution of the NS binary system is not significantly affected by viscous effects. Tidal locking

*Send offprint requests to*: H.-Th. Janka

(synchronization) of NS-NS or NS-BH systems just before tidal disruption requires a dynamic viscosity $\eta \gtrsim 10^{31}\,\mathrm{g\,cm^{-1}s^{-1}}$ (Kochanek 1992, Bildsten & Cutler 1992).

The microscopic shear viscosity of NS matter is much smaller. For matter hotter than the superfluid-transition temperature ($T \gtrsim 10^9\,$K) it is dominated by neutron-neutron interactions. For the density range $10^{14} \lesssim \rho \lesssim 4\cdot 10^{14}\,\mathrm{g/cm^3}$ it was calculated by Flowers & Itoh (1979) and can be fitted to an accuracy of a few percent as $\eta_n = 1.1\cdot 10^{16}\rho_{14}^{9/4}/T_9^2\,\mathrm{g\,cm^{-1}s^{-1}}$ where $\rho_{14} = \rho/10^{14}\,\mathrm{g/cm^3}$ and $T_9 = T/10^9\,$K. For NSs cooler than $10^9\,$K the neutrons become superfluid and their effective scattering cross section with other particles quickly falls to zero. The dynamic shear viscosity is then dominated by electron-electron scattering and is approximated at NS densities to within a few percent by $\eta_e = 6\cdot 10^{16}\,(\rho_{14}/T_9)^2\,\mathrm{g\,cm^{-1}s^{-1}}$ (Cutler & Lindblom 1987). At temperatures of a few $10^9\,$K NSs become opaque for $\nu$'s. Thus shear viscosity as a result of momentum transport due to the scattering of $\nu$'s is potentially important. Assuming as main opacity source the scattering of $\nu$ off $n,p$ (and the charged current absorptions for $\nu_e$ and $\bar\nu_e$), and taking into account the contributions of 3 types of nondegenerate (zero chemical potential) $\nu$'s and $\bar\nu$'s in local thermodynamic equilibrium with the stellar matter, the result of van den Horn & van Weert (1981) for the $\nu$ shear viscosity can be generalized to $\eta_\nu \sim 8.6\cdot 10^{19}T_{10}^2/\rho_{14}\,\mathrm{g\,cm^{-1}s^{-1}}$.

Reisenegger & Goldreich (1994) showed that the energy absorbed by the coalescing NSs is only $\sim 10^{-6}$ of the orbital energy even for stellar oscillation modes that are resonantly excited by the strong tidal forces when the two stars come close. This was confirmed by Lai (1994) who also found that tidal dissipation heats up the stars primarily by resonant excitations of the $g$-modes. Using the microscopic viscosity he found that the core temperature is increased from less than $\sim 10^6\,$K in NSs older than $10^8\,$yr to only $\sim 10^8\,$K before the stars come into contact. To obtain a temperature of about $10^{11}\,$K as estimated by Mészáros & Rees (1992b) would require a viscosity that is larger than the microscopic one by a factor of $10^{10}$–$10^{12}$, corresponding to a dynamic viscosity near the maximum possible value in the NS, $\eta_{\rm max} \approx \rho c R_{\rm ns} \approx 10^{31}\,\mathrm{g\,cm^{-1}s^{-1}}$ ($R_{\rm ns} \approx 10\,$km is the NS radius). Such a high value of the viscosity appears rather unlikely and the NSs should stay comparatively cool before they merge.

*Merging – numerical simulations.* We performed hydrodynamical simulations of the NS merging for three different setups of the initial NS spins. In one case both NS spins were parallel to the orbital angular momentum vector (model B64), meaning rigid rotation of the NS-NS system, in another case both spins were in the opposite direction (C64), and in a third case the NSs had no spins (A64). The two identical, cool Newtonian NSs had a baryonic mass of $M_{\rm ns} \approx 1.63\,M_\odot$ each (corresponding to a gravitational mass of 1.4–1.5 $M_\odot$, dependent on the EOS), a radius of $R_{\rm ns} \approx 15\,$km, and spin periods being the same as the initial orbital period. Initially the stars are spherical and orbit around the common center of mass with a center-to-center distance of 42 km and an orbital period of 2.6 ms. The orbit decays due to gravitational wave emission, dynamical instability sets in, and within $\sim 1.5\,$ms the NSs start to merge.

We chose an initial, central temperature in the stars around $kT \sim 7\,$MeV, corresponding to a thermal energy of roughly 3% of the degeneracy energy for given density $\rho$ and electron fraction $Y_e$. As the stars approach each other, they expand due to tidal stretching. This leads to a slight drop of the core temperatures to about 6 MeV at the centers. These values are close to $10^{11}\,$K and are therefore of the order of the estimates of Mészáros & Rees (1992b) for the final stages of the binary before coalescence. In view of the discussions above we consider them to be far too large. Even if the extreme shear dissipation of a Keplerian disk were present in the coalescing NSs (certainly a strong overestimation of the true shear motions) and no $\nu$ cooling took place, the heating to a temperature of $\sim 6\,$MeV during the last $\sim 24\,$s of the binary spiral-in (starting with cold stars at a separation of $20R_{\rm ns}$) would require a constant viscosity $\eta \sim 10^{26}\,\mathrm{g\,cm^{-1}s^{-1}}$ which is larger than all microscopic viscosities by at least 4 orders of magnitude. The assumed high initial temperatures, however, favor copious and early $\nu$ emission and thus serve the directions of investigation in this work.

During merging the stars heat up drastically because of dynamical compression and dissipation of kinetic energy. In particular at the shear interface where the two stars come into contact first, temperatures between 10 and 30 MeV are obtained, dependent on the NS spins. After $\sim 3\,$ms from the start of the simulations, the two NSs have fused to a compact object with mass $M \approx 3\,M_\odot$, a radius of $\sim 20\,$km, and densities $\rho > (1\,...\,2)\cdot 10^{14}\,\mathrm{g/cm^3}$. The central body reaches temperatures of more than 50 MeV. It is surrounded by a thick disk with orbital velocities of 0.3–0.5$c$, densities $\rho \sim 10^{12}\,\mathrm{g/cm^3}$, and a mass $M_{\rm disk} \approx 0.1$–$0.2\,M_\odot$. During the first few ms after merging this disk is heated from initially 6 MeV to $\sim 10\,$MeV by dissipation and compression in strong pressure waves that are created by the wobbling and oscillations of the central object. Since the EOS used cannot stabilize NSs more massive than $\sim 2.2\,M_\odot$, we expect the massive object to collapse to a BH within a few ms. About 10% of the initial angular momentum are carried away by gravitational waves. With a relativistic rotation parameter $a = Jc/(GM) \lesssim 0.5$–$0.7$ which is smaller than the critical value, $a_{\rm crit} \approx 1$ (Stark & Piran 1985), the central body cannot be stabilized by rotation. Only matter that has a sufficiently large specific angular momentum, $R_{\rm i}v_{\rm k} \lesssim rv \lesssim R_{\rm i}c$, where $R_{\rm i} \approx 6GM/c^2$ is the radius of the innermost stable circular orbit and $v_{\rm k} = \sqrt{GM/R_{\rm i}}$ the Newtonian Keplerian velocity at radius $r = R_{\rm i}$, might remain in a disk around the BH. We estimate that with a typical rotation velocity $v \approx 0.25c$ only gas with orbits $r \gtrsim 44\,...\,107\,$km will have a chance not to be swallowed up by the BH on a dynamical timescale. In our calculations we find that at most 0.01–0.15 $M_\odot$ fulfill this condition, most likely less. This value is extremely sensitive to the angular momentum in the NS-NS system and thus to the assumed NS spins. About 10–30% of the initial angular momentum might be stored in the disk.

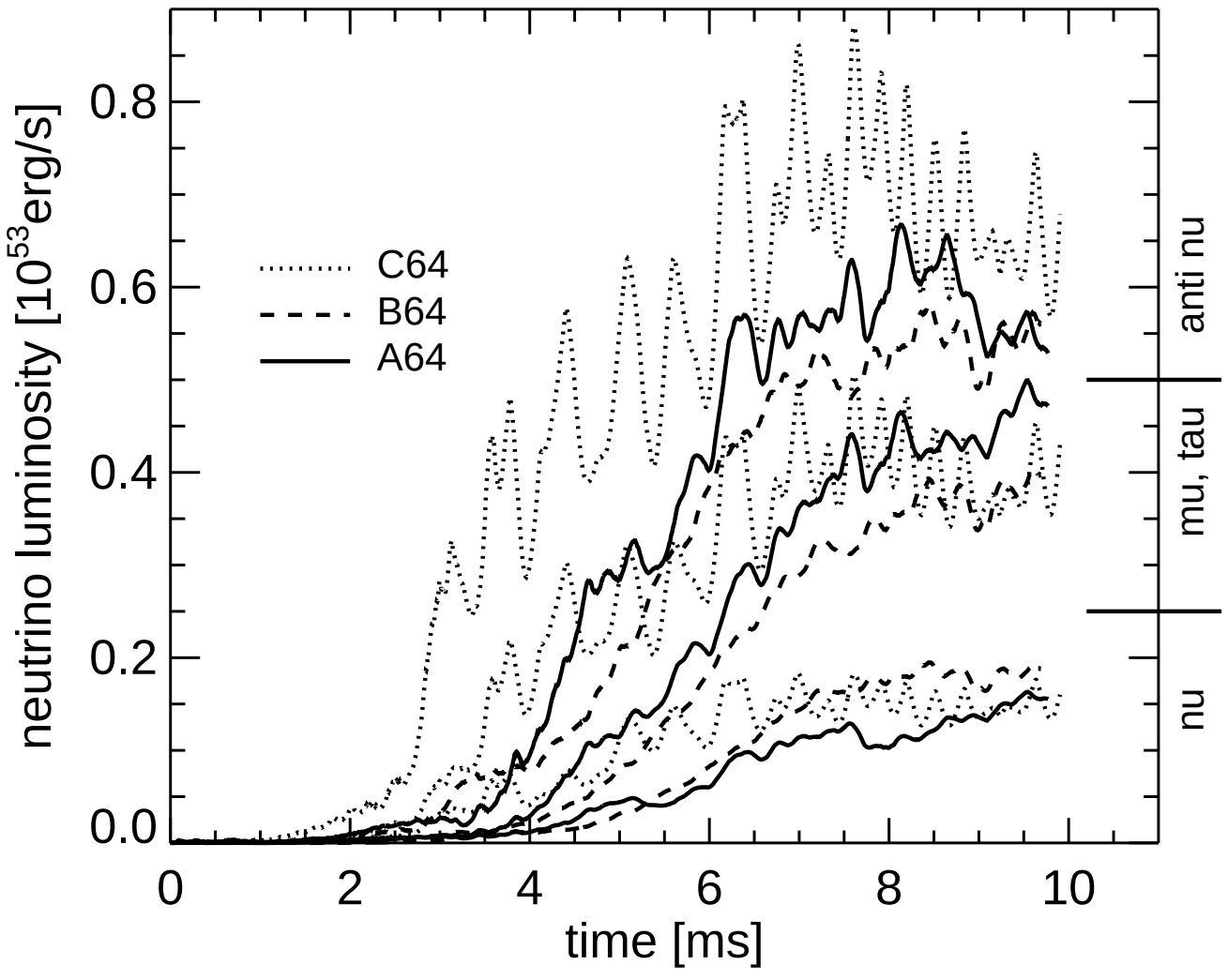

**Fig. 1.** Neutrino fluxes of $\nu_e$, $\bar{\nu}_e$, and $\nu_x \equiv 4\nu_\mu$ for the three models

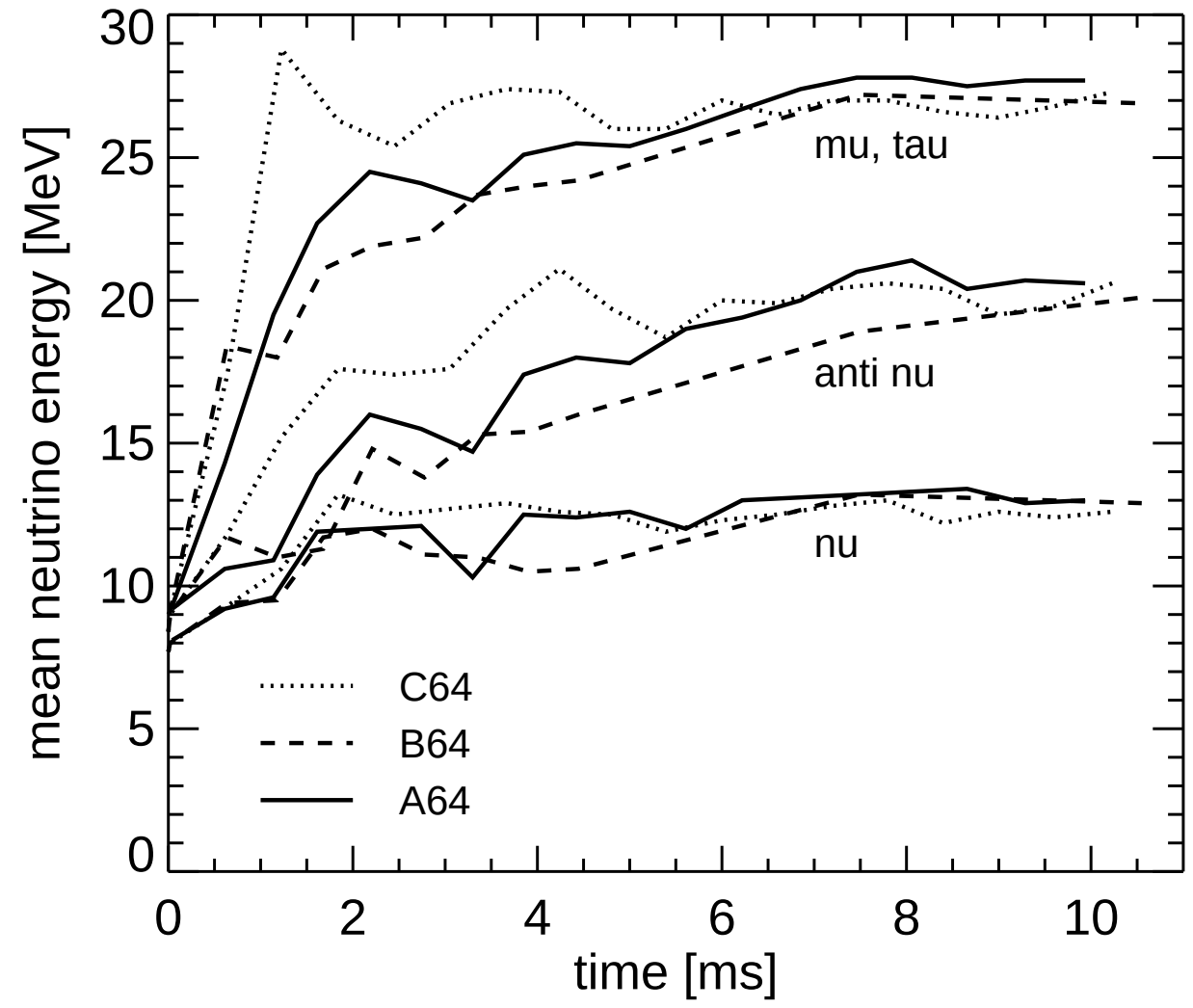

**Fig. 2.** Average energies of emitted $\nu_e$, $\bar{\nu}_e$, and $\nu_x$ for the three models

The luminosities of $\nu_e$, $\bar{\nu}_e$, and $\nu_x$ (the latter representing the sum of the contributions of $\nu_\mu$, $\bar{\nu}_\mu$, $\nu_\tau$, and $\bar{\nu}_\tau$), are plotted in Fig. 1 for all 3 models. The maximum total $\nu$ flux of 1–$1.5 \cdot 10^{53}$ erg/s is reached later than $\sim$ 6 ms and is about 2 orders of magnitude smaller than the peak luminosity in gravitational waves which is obtained during the dynamical merging of the two stars (at $\sim$ 1.5 ms). The total energy emitted in gravitational waves is 0.01–0.02 $M_\odot c^2$, while $\nu$'s radiate away an energy of only 2–3 $\cdot\, 10^{-4}\, M_\odot c^2$ within about 10 ms after the start of the simulations. The corresponding average energies of the emitted $\nu$'s are shown in Fig. 2. The emission is dominated by $\bar{\nu}_e$ with a mean energy of 18–22 MeV. They have a flux of $L_{\bar{\nu}_e} \approx$ (5.5–6.5) $\cdot\, 10^{52}$ erg/s in the stationary situation for $t \gtrsim 6$ ms. The luminosities of $\nu_e$ and $\nu_x$ are smaller, $L_{\nu_e} : L_{\bar{\nu}_e} : 4L_{\nu_\mu} \approx 1 : 4 : 3$, and their mean energies are $\langle\epsilon_{\nu_e}\rangle \approx$ 12–13 MeV and $\langle\epsilon_{\nu_x}\rangle \approx$ 26–28 MeV, respectively. By far most of the $\nu$ emission comes from the disk. After having reached its maximum interior temperatures $kT \approx 10$ MeV at $t \gtrsim 5$ ms, the disk emits more than 90% of the total flux. While the very dense, much more opaque core loses energy by $\nu$ emission at a typical rate of $\sim 10^{31}$–$10^{32}$ erg cm$^{-3}$s$^{-1}$, the disk emits $\nu$'s at a rate of (3–10) $\cdot\, 10^{32}$ erg cm$^{-3}$s$^{-1}$. When decompressed and heated, $n$-rich ($Y_e \sim 0.01$–$0.09$) NS matter becomes moderately degenerate and the $\beta$-equilibrium is shifted to $\mu_e + \mu_p - \mu_n = \mu_{\nu_e} < 0$. In such a situation the process $e^+ + n \to p + \bar{\nu}_e$ is faster than $e^- + p \to n + \nu_e$ and leads to the dominant production of $\bar{\nu}_e$ in the disk.

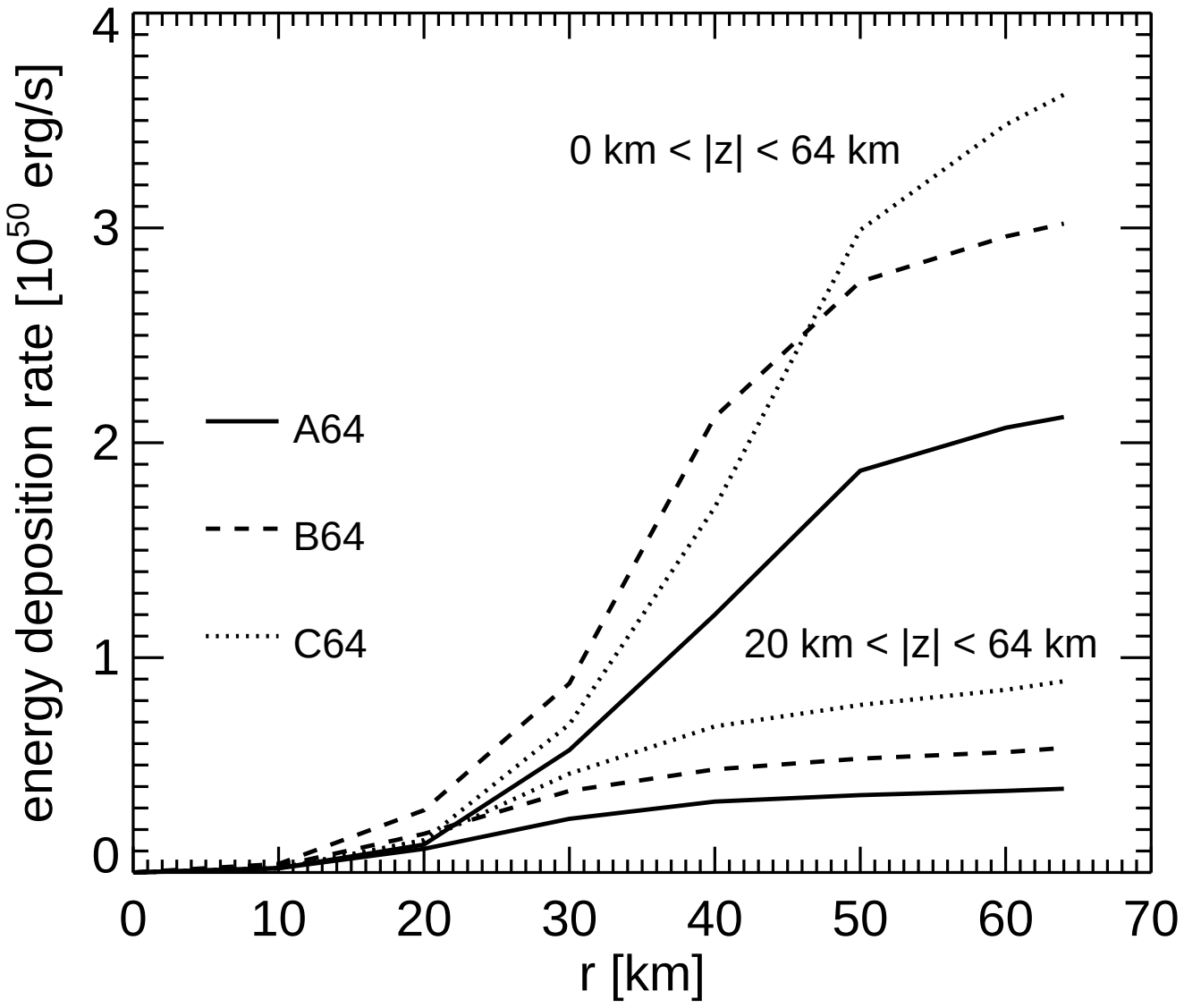

**Fig. 3.** Integrated $\nu\bar{\nu}$ energy deposition rate in the region around the merged neutron stars out to an equatorial radius $r$ in the vertical range $0 < |z| < 64$ km and in the range $20\,\mathrm{km} < |z| < 64$ km, respectively

Our models were evaluated for the local energy deposition rates by $\nu\bar{\nu}$ annihilation in the vicinity of the merged NSs in regions where $\rho < 10^{11}$ g/cm$^3$ and where the energy loss rate by $\nu$ emission is less than $10^{30}$ erg s$^{-1}$cm$^{-3}$ (for details see Ruffert et al., in preparation). Fig. 3 displays the cumulative energy deposition rate $\mathrm{d}E_{\nu\bar{\nu}}(r)/\mathrm{d}t$, integrated out to a radius $r$ in the equatorial plane ($r \leq 64$ km), once for the vertical range $|z| \leq 64$ km, and another time only for the volume that lies above and below the thick disk around the central object ($20\,\mathrm{km} \leq |z| \leq 64$ km). For the disk geometry of our models we find with the numbers of Figs. 1, 2, and 3 for the efficiency $e_{\nu\bar{\nu}} \equiv (\mathrm{d}E_{\nu\bar{\nu}}/\mathrm{d}t)/(L_{\nu_e} + L_{\bar{\nu}_e} + 4L_{\nu_\mu})$:

$$e_{\nu\bar{\nu}} \lesssim (2\ldots3)\cdot 10^{-3}\, \frac{L_{\nu_e}}{1.5\cdot 10^{52}\mathrm{erg/s}}\, \frac{\langle\epsilon_{\nu_e}\rangle}{13\,\mathrm{MeV}}\, \frac{20\,\mathrm{km}}{R_{\mathrm{disk}}}\,, \qquad (1)$$

where $R_{\mathrm{disk}}$ is the inner radius of the disk. However, 75–80% of the annihilation energy is deposited in or near the disk (Fig. 3) and will serve to drive a baryonic wind. The fraction of the $\nu$ energy going into $e^+e^-$ pairs in a baryon-free funnel along the $z$-axis and thus being potentially relevant to produce energetic $\gamma$-rays is at most only $\sim (0.4\text{–}0.9)\cdot 10^{50}$ erg/s at the time of maximum luminosity ($t \gtrsim 6$ ms).

*Post-merging evolution.* Our simulations suggest that some material, possibly $\sim 0.1\,M_\odot$, could remain in a disk around the central BH. The energy for a $\gamma$-ray burst might be provided by the $\nu$ and $\bar\nu$ emitted from this disk. The efficiency of $\nu\bar\nu$ annihilation increases linearly with the $\nu$ luminosity (Eq. (1)). The $\nu$ emission properties of the disk should be discussed self-consistently with the disk evolution for which the viscous momentum transport and viscous heating plays a crucial role. The duration of $\nu$ emission, accretor mass, disk radius, and disk temperature cannot be chosen and varied independently. We now try to relate these quantities by simple conservation arguments for disk and $\nu$ emission.

The lifetime of the disk will decrease with larger viscosity $\eta$ because the viscous force that generates a torque carrying angular momentum outwards is increased. Setting the rate of change of angular momentum equal to the torque exerted by the viscous stress yields an estimate for the accretion rate of disk material: $\mathrm{d}M/\mathrm{d}t \sim 6\pi\eta h$, where $2h \sim 2R_\mathrm{s}$ is the vertical diameter of the thick disk around the BH with mass $M$ and Schwarzschild radius $R_\mathrm{s} \approx 9(M/3\,M_\odot)$ km. This leads to an accretion timescale $t_\mathrm{acc} \sim \Delta M_\mathrm{d}/(6\pi\eta R_\mathrm{s})$ for a disk mass $\Delta M_\mathrm{d}$. For small $\eta$, the disk will stay cool and nearly transparent to $\nu$'s. The total $\nu$ flux $L_\nu$ will therefore increase with the viscous energy dissipation rate. The latter is approximated by $\mathrm{d}Q/\mathrm{d}t \sim \frac{9}{4}\eta GM/R_\mathrm{d}^3$ for a (Newtonian) Keplerian disk with an inner radius $R_\mathrm{d} \sim 3R_\mathrm{s} \sim 6GM/c^2$ that is taken to be the innermost stable circular orbit around the central accreting BH. The timescale of $\nu$ emission by the disk scales like $t_\mathrm{acc} \propto 1/\eta$ and the $\nu$ luminosity like $L_\nu \propto \eta$ in the optically thin case. The energy converted into $e^+e^-$ by $\nu\bar\nu$ annihilation is therefore (Eq. (1)) $E_{\nu\bar\nu} \propto L_\nu^2 t_\mathrm{acc} \propto \eta$. If the disk viscosity $\eta$ is large, the disk will be strongly heated and become very opaque against $\nu$'s. The $\nu$ luminosity then behaves like $L_\nu \sim V_\mathrm{d}(\mathrm{d}Q/\mathrm{d}t)t_\mathrm{acc}/t_\mathrm{diff}$ for a disk volume $V_\mathrm{d}$, and $E_{\nu\bar\nu}$ decreases with $\eta$ according to $E_{\nu\bar\nu} \propto 1/(\eta t_\mathrm{diff}^2)$.

The dependence of $E_{\nu\bar\nu}$ on $\eta$ in the $\nu$-transparent and optically thick cases suggests that there is a maximum of the annihilation energy for that value of $\eta$ at which $t_\mathrm{acc} \approx t_\mathrm{diff}$. Assuming the part of the disk where most of the $\nu$'s are emitted to have mass $\Delta M_\mathrm{d}$ and to be a (homogeneous) torus with center at $4R_\mathrm{s}$ and radius $R_\mathrm{s}$ (Mochkovitch et al. 1993, Jaroszyński 1993), and using $t_\mathrm{diff} \sim 3R_\mathrm{s}^2\rho\sigma/(m_\mathrm{u}c)$ with the $\nu$ interaction cross section $\sigma$ and the atomic mass unit $m_\mathrm{u}$, one determines the value of the disk viscosity where $\nu\bar\nu$ annihilation yields the largest energy as $\eta^* \sim \frac{4}{9}\pi m_\mathrm{u}c/\sigma$. With $\sigma \sim 4.3\cdot 10^{-41}(kT/5\,\mathrm{MeV})^2$ cm$^2$ this gives $\eta^* \sim 1.6\cdot 10^{27}(kT/5\,\mathrm{MeV})^{-2}$ g cm$^{-1}$s$^{-1}$.

Considering a disk with this optimum value $\eta^*$, one can estimate the internal temperature $T_\mathrm{int}$ of the disk from equating $L_\nu \sim V_\mathrm{d}(\mathrm{d}Q/\mathrm{d}t)$ with $L_\nu \sim \varepsilon_\nu V_\mathrm{d}/t_\mathrm{diff}$. Here $\varepsilon_\nu \sim 3\cdot\frac{7}{8}a_\mathrm{rad}T_\mathrm{int}^4$ is the sum of the energy densities of all 3 kinds of non-degenerate $\nu\bar\nu$ pairs with $a_\mathrm{rad}$ being the radiation constant. One finds $(kT_\mathrm{int}/5\,\mathrm{MeV}) \approx 10(\Delta M_\mathrm{d,01})^{1/4}R_\mathrm{s,9}^{-3/4}$ where $\Delta M_\mathrm{d,01} \equiv \Delta M_\mathrm{d}/(0.1\,M_\odot)$ and $R_\mathrm{s,9} \equiv R_\mathrm{s}/(9\,\mathrm{km})$ is normalized to the Schwarzschild radius of a $3\,M_\odot$ BH. Using this and the requirement that $L_\nu \sim S_\mathrm{d}(3\frac{7}{8}a_\mathrm{rad}T_\mathrm{surf}^4)c/4$ ($S_\mathrm{d}$ is the surface of the toroidal $\nu$ emitting disk) allows one to determine the surface temperature: $(kT_\mathrm{surf}/5\,\mathrm{MeV}) \approx 0.7(\Delta M_\mathrm{d,01})^{-1/8}R_\mathrm{s,9}^{1/8}$. With $kT_\mathrm{int} \approx 50$ MeV the optimum value of the disk viscosity becomes $\eta^* \sim 1.6\cdot 10^{25}$ g cm$^{-1}$s$^{-1}$. This corresponds to an effective $\alpha$-parameter $\alpha \approx \eta^*/(\rho c_\mathrm{s}R_\mathrm{s}) \sim (3\text{–}6)\cdot 10^{-4}$, $c_\mathrm{s}$ being the sound speed and the range of $\alpha$ values accounting for differences when the matter is relativistic or non-relativistic.

For the diffusion and accretion timescales one obtains $t_\mathrm{acc} \approx t_\mathrm{diff} \sim 0.67\,(\eta^*)^{-1}(\Delta M_\mathrm{d,01})R_\mathrm{s,9}^{-1}$ s. This suggests that the disk is optically thick for $\nu$'s. Under the described conditions the dynamical shear viscosity due to $\nu$'s is $\eta_\nu \sim 0.9\cdot 10^{25}(kT_\mathrm{int}/50\,\mathrm{MeV})^2(\Delta M_\mathrm{d,01})^{-1}R_\mathrm{s,9}^3$ g cm$^{-1}$s$^{-1}$ which, interestingly, is of the right size to account for $\eta^*$. The $\nu$ luminosity is $L_\nu \sim 5\cdot 10^{52}(\Delta M_\mathrm{d,01})^{-1/2}R_\mathrm{s,9}^{5/2}$ erg/s. The total energy radiated away over the time $t_\mathrm{acc}$ becomes $L_\nu t_\mathrm{acc} \sim 3.4\cdot 10^{52}\Delta M_\mathrm{d,01}$ erg which is equal to the gravitational binding energy of mass $\Delta M_\mathrm{d}$ when it is swallowed by the BH at $r = 3R_\mathrm{s}$: $E_\mathrm{grav} \approx GM\Delta M_\mathrm{d}/(3R_\mathrm{s}) = \frac{1}{6}\Delta M_\mathrm{d}c^2$. With the efficiency of Eq. (1), using $L_{\nu_e} \sim (\frac{1}{7}\ldots\frac{1}{5})L_\nu \sim 10^{52}$ erg/s, $\langle\epsilon_{\nu_e}\rangle \approx 3kT_\mathrm{surf} \sim 11$ MeV, and $R_\mathrm{disk} = 3R_\mathrm{s}$, we find for the total energy transformed into $e^+e^-$ pairs by annihilation of $\nu$ and $\bar\nu$ radiated from the disk:

$$E_{\nu\bar\nu} \sim (2\ldots 4)\cdot 10^{49}\,(\Delta M_\mathrm{d,01})^{3/8}\,R_\mathrm{s,9}^{13/8}\ \mathrm{erg}\,. \qquad (2)$$

This result was derived for the most favorable conditions to get large values for $E_{\nu\bar\nu}$. In particular, it was assumed that the dynamical viscosity of the disk attains the optimum value $\eta^*$. Also, the efficiency $e_{\nu\bar\nu}$ of Eq. (1) most likely overestimates the useful fraction of the annihilation energy by a considerable factor. From our hydrodynamical simulations we, moreover, deduce that a disk with a mass $\gtrsim 0.1\,M_\odot$ will be formed in NS-NS mergers only under very special conditions, e.g., when there is high angular momentum in the system due to NS spin like in our model C64. The smaller angular momentum in models A64 and B64 allows much less material to remain in a disk. These arguments might also be true for NS-BH mergers. Finally, general relativistic effects were completely neglected in the presented considerations. Jaroszyński (1993) showed that they significantly lower the energy that can be transported to infinity.

Therefore it seems to be extremely hard to account for the energies of cosmological $\gamma$-ray bursts by $\nu\bar\nu$ annihilation. The same conclusion was arrived at on grounds of different arguments by Jaroszyński (1995) who found that only BH's with a high angular momentum ($a \sim 1$) yield promising models. This favors collapsed cores of rapidly rotating WR stars in the "failed supernova" scenario (Woosley 1993). Certainly, our analytic considerations do not include effects due to possible convective overturn or instabilities in the disk which might increase the $\nu$ fluxes and the average $\nu$ energies considerably and thus might help $\nu\bar\nu$ annihilation. Also, magnetic fields in the merging NSs and in the disk were neglected. Within a lifetime of several tenths of a second initial $B$-fields of $\sim 10^{12}$–$10^{13}$ G might be amplified by factors of a few 100 in the rapidly rotating disk around the BH (rotation periods $\sim 1$ ms) and might become energetically important (Rees, personal communication).